# "Measuring and assessing economic uncertainty"

Oscar Claveria

## Abstract

This paper evaluates the dynamic response of economic activity to shocks in agents' perception of uncertainty. The study focuses on the comparison between manufacturers 'and consumers' perception of economic uncertainty. Since uncertainty is not directly observable, we approximate it using the geometric discrepancy indicator of Claveria et al. (2019). This approach allows us quantifying the proportion of disagreement in business and consumer expectations of eleven European countries and the Euro Area. First, we compute three independent indices of discrepancy corresponding to three dimensions of uncertainty (economic, inflation and employment) and we average them to obtain aggregate disagreement measures for businesses and for consumers.

Next, we use a bivariate Bayesian vector autoregressive framework to estimate the impulse response functions to innovations in disagreement in every country. We find that the effect on economic activity of shocks to the perception of uncertainty differ markedly between manufacturers and consumers. On the one hand, shocks to consumer discrepancy tend to be of greater magnitude and duration than those to manufacturer discrepancy. On the other hand, innovations in disagreement between the two collectives have an opposite effect on economic activity: shocks to manufacturer discrepancy lead to a decrease in economic activity, as opposed to shocks to consumer discrepancy. This finding is of particular relevance to researchers when using cross-sectional dispersion of survey-based expectations, since the effect on economic growth of shocks to disagreement depend on the type of agent.

Oscar Claveria: AQR-IREA, Department of Econometrics, Statistics and Applied Economics, University of Barcelona, 08034 Barcelona, Spain. Email: oclaveria@ub.edu

# 1. Introduction

The analysis of economic uncertainty has gained renewed interest since the advent of the 2008 financial crisis. While there is a widespread consensus that uncertainty shocks have an effect on real activity (Bachmann and Bayer 2013; Baker et al. 2016; Bloom 2009; Paloviita and Viren 2014; Zarnowitz and Lambros 1987), there are several strategies to measure uncertainty. Since economic uncertainty is not directly observable, some authors have opted to proxy it by using the realized volatility in equity markets (Basu and Bundick 2012; Bekaert et al. 2013; Caggiano et al. 2014; Yıldırım-Karaman 2017) or in oil and natural gas prices (Atalla et al. 2016; Hailemariam and Smyth 2019). Other authors have used econometric unpredictability, understood as the conditional volatility of the unforecastable components of a broad set of economic variables (Chuliá et al. 2017; Jurado et al. 2015; Meinen and Roehe 2017). The ex-post nature of this latter approach, has recently generated a strand of the empirical research that makes use of survey-derived measures of expectations dispersion (Binder 2017; Binding and Dibiasi 2017; Clements and Galvão 2017; Krüger and Nolte 2016).

Disagreement measures based on survey expectations make use of prospective information, as agents are asked about the expected future evolution of a wide range of variables. The ex-ante nature of survey expectations makes them especially appropriate to evaluate the anticipatory properties of disagreement-based uncertainty indicators. While most studies rely on quantitative macroeconomic expectations made by professional forecasters (Dovern 2015; Lahiri and Sheng 2010; Mankiw et al. 2004; Oinonen and Paloviita 2017), an alternative source of survey expectations are business and consumer tendency surveys (Bachmann et al. 2013; Claveria 2020; Girardi and Reuter 2017; Meinen and Roehe 2017; Mitchell et al. 2007; Mokinski et al. 2015).

The European Commission conducts monthly business and consumer tendency surveys in which respondents are asked whether they expect a set of variables to rise, fall or remain unchanged. Firms are asked about production, selling prices, employment and other variables concerning developments in their sector, and households are asked about their spending intentions and the general economic situation influencing those decisions. We use the information coming from both surveys to elicit agents' expectations about production and economic activity, prices, and employment in eleven European countries and the Euro Area (EA): Austria, Belgium, Finland, France, Germany, Greece, Italy, the Netherlands (NL), Portugal, Spain, and the United Kingdom (UK).

In this research we use qualitative survey data from two independent tendency surveys conducted by the European Commission, the industry survey and the consumer survey. This dual approach allows us to simultaneously measure disagreement about economic activity, prices and employment in both business and consumer expectations. By means of Claveria et al.'s (2019) geometric indicator of discrepancy, we proxy the three different dimensions of uncertainty, which we then use to construct aggregate disagreement indicators for both businesses (*DB*) and consumers (*DC*).

We apply a bivariate Bayesian vector autoregressive (BVAR) framework to analyse the dynamic response of economic growth to innovations in each type of disagreement: manufacturers' vs consumers'. This study contributes to the existing literature by providing a comparative view of firms versus consumers of the dynamic relationship between the perception of economic uncertainty and the evolution of economic activity.

The paper is organised as follows. The next section introduces the data and the methodological approach. Empirical results are provided in Section 3. Finally, concluding remarks and future lines of research are drawn in Section 4.

## 2. Data and Methodology

2.1. Data

The empirical analysis focuses on manufacturing firms' and consumers' expectations about the future evolution of economic activity, inflation and unemployment. We use monthly data from the joint harmonised EU industry and consumer surveys conducted by the European Commission. Regarding the quantitative information, we use annual rates of change of the gross domestic product (GDP) provided by Eurostat. The sample period goes from May 2005 to December 2017.

In the survey, manufacturers are asked about their expectations regarding production, selling prices and employment for the months ahead, and they are faced with three options: "up", "unchanged" and "down". The aggregated percentages of the individual replies in each category are respectively denoted as $P_t$, $E_t$, and $M_t$.

Consumers, for their part, are asked how they think the general economic situation, the cost of living, and the level of unemployment in the country will change over the next twelve months. Consumers have three additional response categories: two at each end ("a lot better/much higher/sharp increase", and "a lot worse/much lower/sharp decrease"), and a "don't know" option. We opt for grouping all positive responses in *P*, all negative ones in *M*, and incorporating the "don't know" share in *E* for each time period.

2.2 Measurement of uncertainty

The most common way of presenting survey results is the balance, obtained as $P_t - M_t$. The most widespread measures of disagreement among survey respondents use the dispersion of balances as a proxy for uncertainty (Bachmann et al. 2013; Girardi and Reuter 2017). Bachmann et al. (2013) proposed an indicator of disagreement based on the square root of the variance of the balance:

$$DISP_t = \sqrt{P_t + M_t - (P_t - M_t)^2} \tag{1}$$

The omission of the information contained in the "no change" category led Claveria et al. (2019) to develop a disagreement metric that incorporated the information coming from all the reply options, whose number is denoted as *N*. Given that the sum of the shares of responses adds to a hundred, the authors compute an *N*-dimensional vector that aggregates the information from all answering categories and project it as a point on a simplex of $N-1$ dimensions that encompasses all possible combinations of responses. For $N=3$, the simplex takes the form of an equilateral triangle (Fig. 1), where the point *V* corresponds to a unique convex combination of the three reply options for each period in time. See Claveria (2018) for an extension of the methodology for a larger number of reply options, and Claveria (2019) for an application of the methodology when $N=5$.

Insomuch as all vertices are at the same distance to the centre of the simplex (*O*), the ratio of the distance of a point to the barycentre (*VO*) and the distance from the barycentre to the nearest vertex (*OP*) provides the proportion of agreement among respondents. Consequently, the indicator of discrepancy for a given period in time can be formalised as:

$$D_t = 1 - \left[ \frac{\sqrt{(P_t - 1/3)^2 + (E_t - 1/3)^2 + (M_t - 1/3)^2}}{\sqrt{2/3}} \right] \tag{2}$$

This metric is bounded between zero and one, and conveys a geometric interpretation. The center of the simplex corresponds to the point of maximum disagreement, indicating that the answers are equidistributed among the three response categories. Conversely, each of the $N$ vertexes corresponds to a point of minimum disagreement, where one category draws all the answers and $D_t$ reaches the value of zero.

**Fig. 1.** Projection of the combination of the three reply options

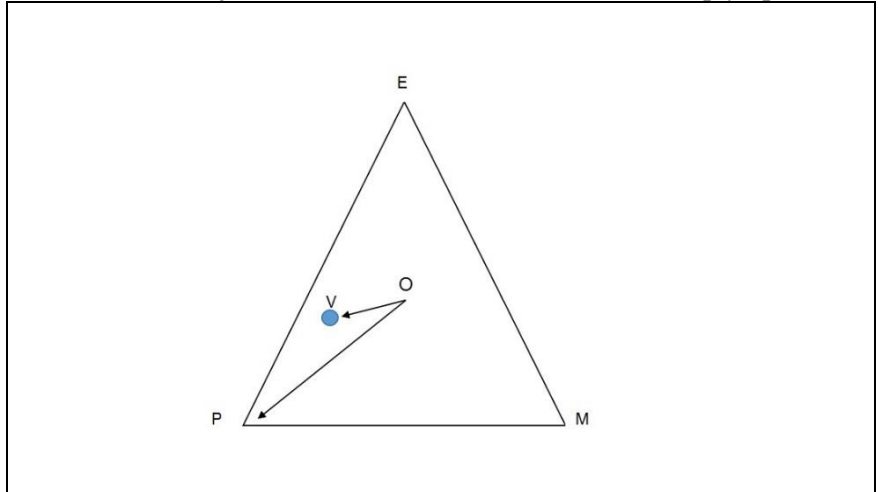

Notes. $V$ is the vector of the three aggregated reply options for a given period in time: $P$ corresponds to the % of "increase" replies, $M$ to the % of "fall", and $E$ to the % of "remains constant". $O$ represents the centre of the simplex (barycentre), which corresponds to the point of maximum disagreement.

In Fig. 2 and Fig. 3 we compare the evolution of the geometric measure of disagreement (2) to that of the standard deviation of the balance (1) in the EA for the question regarding firms' expectations about future production (Fig. 2), and for households' expectations about the general economic situation (Fig. 3). We can observe that both series co-evolve. Claveria (2020) obtained a high positive correlation between measures (1) and (2) of disagreement, and found that the main difference between both measures mainly lied in their average level and dispersion, being *DISP* more volatile and higher in most countries. By means of a simulation experiment, Claveria et al. (2019) showed that the omission of neutral responses in (1) resulted in an overestimation of the level of disagreement.

In this study we apply expression (2) to measure discrepancy in manufacturing surveys ($DB_t$) and in consumer surveys ($DC_t$). Table 1 contains the summary statistics of disagreement in business and consumer surveys. For all countries except Greece, the average degree of *DC* is higher than *DB*. We also observe notable differences between *DB* and *DC* in some countries. In this sense, Belgium shows the lowest average DB level and the highest average DC level, similarly to the Netherlands and Germany to a lesser extent.

**Fig. 2** Evolution of disagreement measures for firms' expectations about industrial production in the EA (2005.05-2017.12)

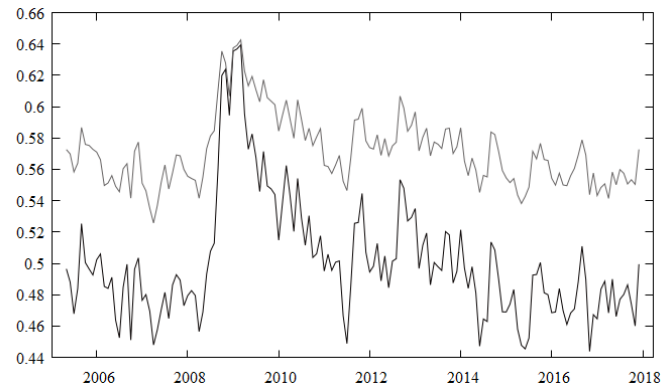

Notes. The solid darker black line represents the evolution of the geometric measure of disagreement (*D*), while the clearer black line represents the evolution of the standard deviation of the balance statistic (*DISP*).

**Fig. 3** Evolution of disagreement measures for households' expectations about the general economic situation in the EA (2005.05-2017.12)

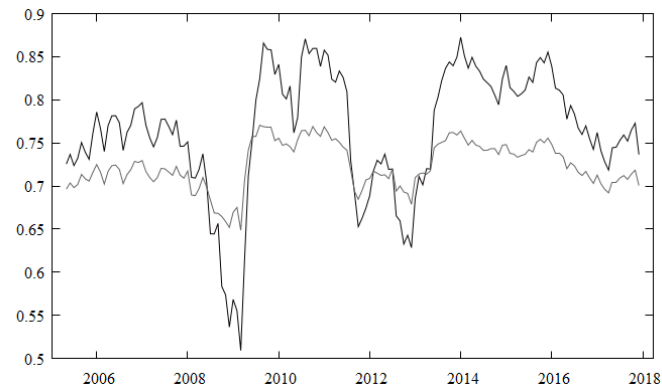

Notes. The solid darker black line represents the evolution of the geometric measure of disagreement (*D*), while the clearer black line represents the evolution of the standard deviation of the balance statistic (*DISP*).

Regarding the correlation with GDP growth, we observe that DB shows a negative correlation in all countries except the UK, while DC shows positive correlations with economic growth in all cases. Greece, Spain and Belgium are the economies where we obtain the highest correlations between DB and economic growth dynamics. Greece is also the country in which we obtain the highest correlation between DC and GDP growth. We want to note that, to some extent, the discrepancies between firms and consumers can be partly attributable to differences in the questions in both surveys: while consumer survey questions refer to objective variables, business surveys questions refer to firm-specific factors.

**Table 1**
Descriptive and correlation analysis – Disagreement and GDP growth

|  | DB | | DC | |
|---|---|---|---|---|
|  | Mean | SD | Mean | SD |
| Austria | 0.447 | 0.034 | 0.634 | 0.028 |
| Belgium | 0.371 | 0.071 | 0.731 | 0.111 |
| Finland | 0.539 | 0.034 | 0.601 | 0.023 |
| France | 0.533 | 0.032 | 0.585 | 0.041 |
| Germany | 0.398 | 0.042 | 0.587 | 0.031 |
| Greece | 0.501 | 0.042 | 0.459 | 0.087 |
| Italy | 0.425 | 0.025 | 0.553 | 0.055 |
| Netherlands | 0.377 | 0.051 | 0.671 | 0.055 |
| Portugal | 0.432 | 0.061 | 0.535 | 0.079 |
| Spain | 0.421 | 0.040 | 0.608 | 0.072 |
| United Kingdom | 0.607 | 0.043 | 0.684 | 0.062 |
| Euro Area | 0.443 | 0.033 | 0.646 | 0.044 |
|  | Correlation | | Correlation | |
|  | GDP growth and DB | | GDP growth and DC | |
| Austria | -0.323 | | 0.492 | |
| Belgium | -0.659 | | 0.295 | |
| Finland | 0.131 | | 0.453 | |
| France | -0.587 | | 0.305 | |
| Germany | -0.391 | | 0.584 | |
| Greece | -0.728 | | 0.773 | |
| Italy | -0.251 | | 0.284 | |
| Netherlands | -0.641 | | 0.613 | |
| Portugal | -0.547 | | 0.569 | |
| Spain | -0.711 | | 0.235 | |
| United Kingdom | 0.106 | | 0.495 | |
| Euro Area | -0.601 | | 0.532 | |

Notes: SD denotes standard deviation. DB refers to aggregate disagreement for businesses and DC to aggregate disagreement for consumers.

In Fig. 4 we compare the evolution of DB to that of DC in each country. We observe a strong a negative correlation between both measures. To further explore the linear degree of association between *DB* and *DC*, in Fig. 5 we show the cross-correlograms between *DB* and lagged *DC*. We corroborate that there is a contemporaneous negative correlation between the measures of disagreement of both collectives in all countries. The highest value is obtained for Portugal, followed by the EA, Spain, Greece, the Netherlands and Belgium. The lowest values are found in Finland and Germany, where there is almost no association between *DB* and *DC*.

**Fig. 4.** Business disagreement vs Consumer disagreement

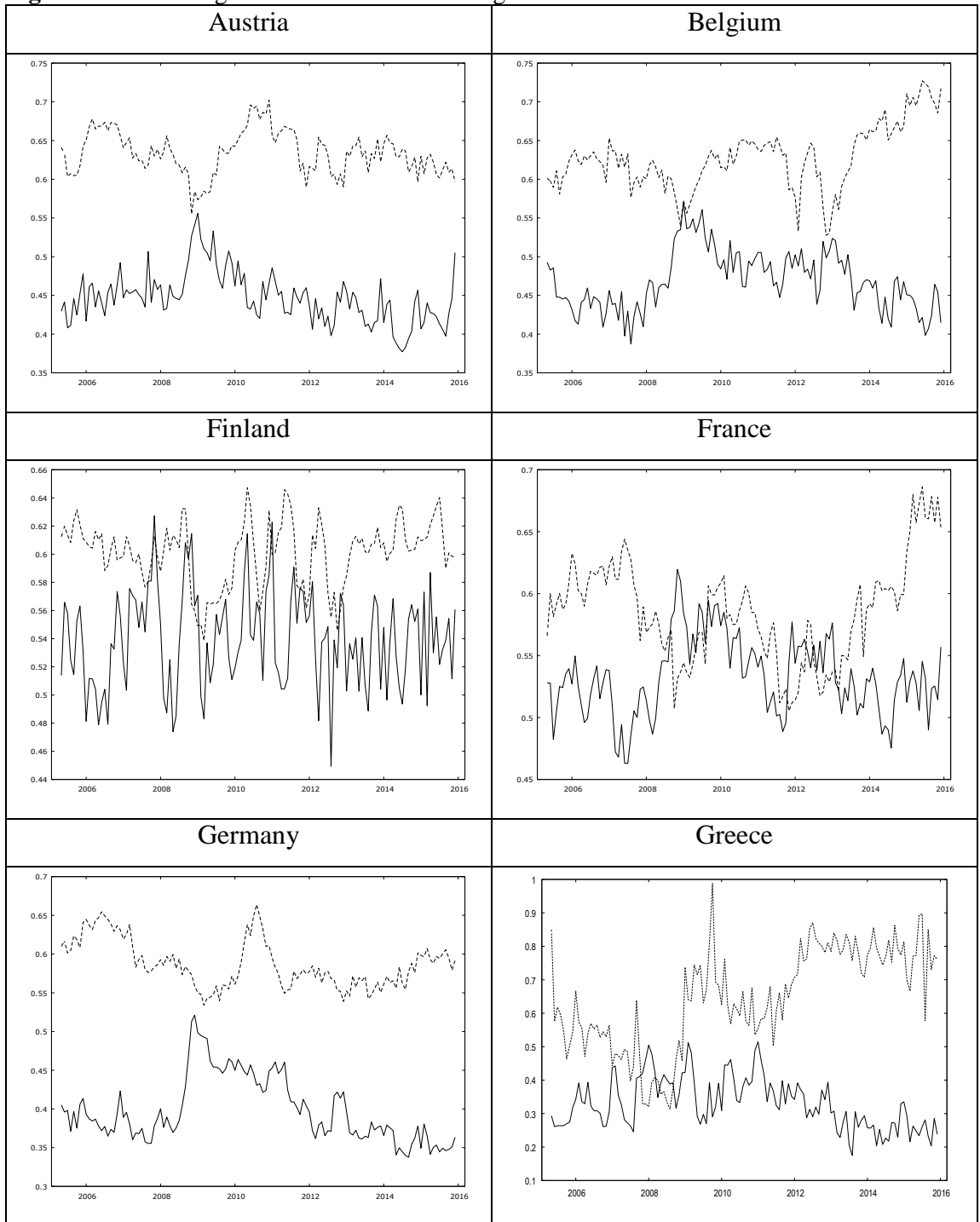

Notes: The solid black line represents the evolution of the geometric measure of manufacturing disagreement – aggregate disagreement for businesses (*DB*), while the dotted black line represents the evolution of aggregate disagreement for consumers (*DC*).

**Fig. 4** (cont.)**.** Business disagreement vs Consumer disagreement

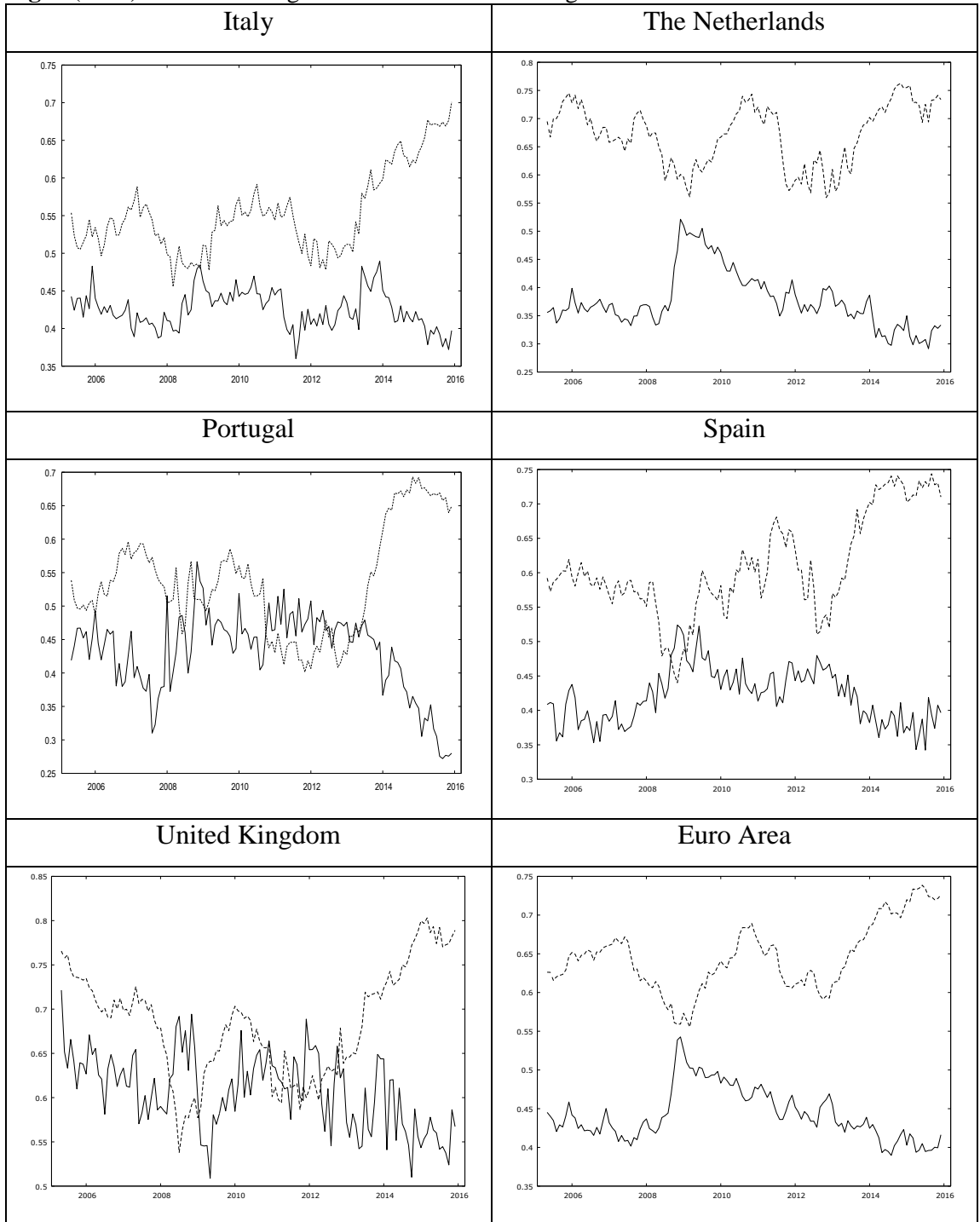

Notes: The solid black line represents the evolution of the geometric measure of manufacturing disagreement – aggregate disagreement for businesses (*DB*), while the dotted black line represents the evolution of aggregate disagreement for consumers (*DC*).

**Fig. 5.** Cross-correlograms – *DB* vs lagged *DC*

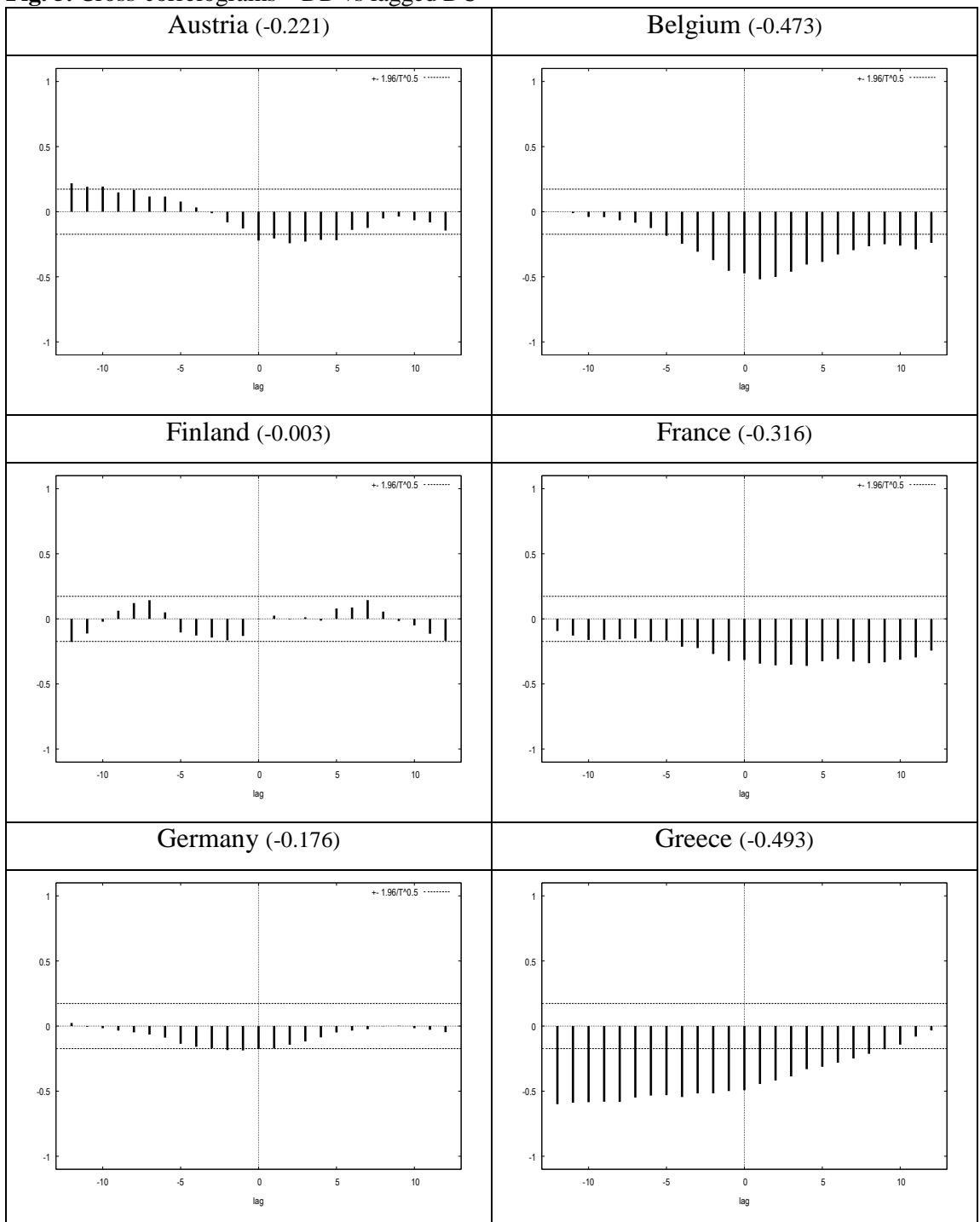

Notes: The abscissa axis represents the number of lags/leads in aggregate disagreement of consumers (*DC*). Contemporaneous correlation between brackets.

**Fig. 5** (cont.)**.** Business disagreement vs Consumer disagreement

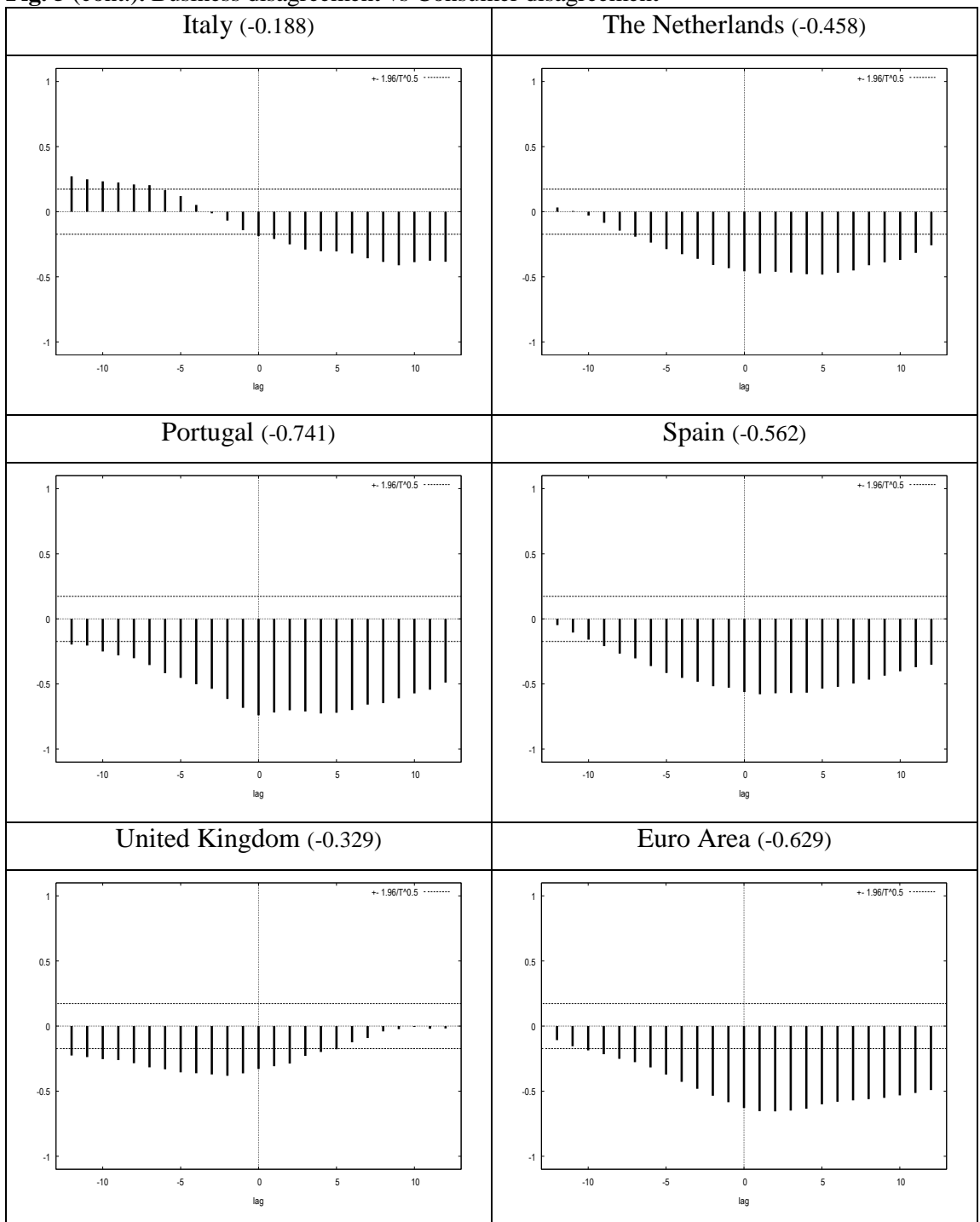

Notes: The abscissa axis represents the number of lags/leads in aggregate disagreement of consumers (*DC*). Contemporaneous correlation between brackets.

## 3. Empirical results

There exists empirical evidence on the bidirectional relationship between uncertainty and macroeconomic variables (Alessandri and Mumtaz 2019; Glocker and Hölzl 2019; Gupta et al. 2019; Mumtaz and Musso 2019). By means of a VAR approach, in this section we first examine the dynamic relationship of the discrepancy measures computed in the previous section to gauge the perception of uncertainty and the corresponding macromagnitudes. As we are estimating independent vector autoregressions per country and no spillover effects are considered, we introduce an index $i = 1, \ldots, N$ to denote the $N$ countries analysed in the study. We use the following bivariate model per country:

$$x_{it} = \sum_{p=1}^{P} A_{ip} x_{it-p} + \varepsilon_{it}, \quad \varepsilon_{it} \sim N(0, \Sigma_i) \tag{3}$$

With $x_{it} = (D_{\bullet,it}, z_{it})$, where $D_{\bullet,it}$ refers to the proposed disagreement measure for businesses (*DB*) and consumers (*DC*) respectively and, $z_{it}$ refers to the macroeconomic variable of reference, which in our case is output growth for the *i*-th country at time *t* $(t = 1, \ldots, T)$. The number of lags, *p*, is selected by means of Schwarz's Bayesian information criterion (BIC). We use heteroscedasticity-consistent standard errors for the estimation. Thus, in the resulting two-variable VAR models each of the uncertainty measures (*DB* and *DC*) is related to GDP growth.

We estimate independent vector autoregressions per country, so no spillover effects are considered. We have used a Cholesky decomposition of the covariance matrix ordering the uncertainty proxies first (Bloom, 2009). We have applied a Bayesian estimation procedure, using the Minnesota type prior to achieve shrinkage and more precise inference. We have replicated the analysis for each country and the Euro Area.

In Fig. 6 we compare the estimated impulse response functions (IRFs) of output growth to innovations in manufacturers' and consumers' perception of uncertainty as captured by the discrepancy measures.

**Fig. 6.** IRFs of GDP to shocks in disagreement

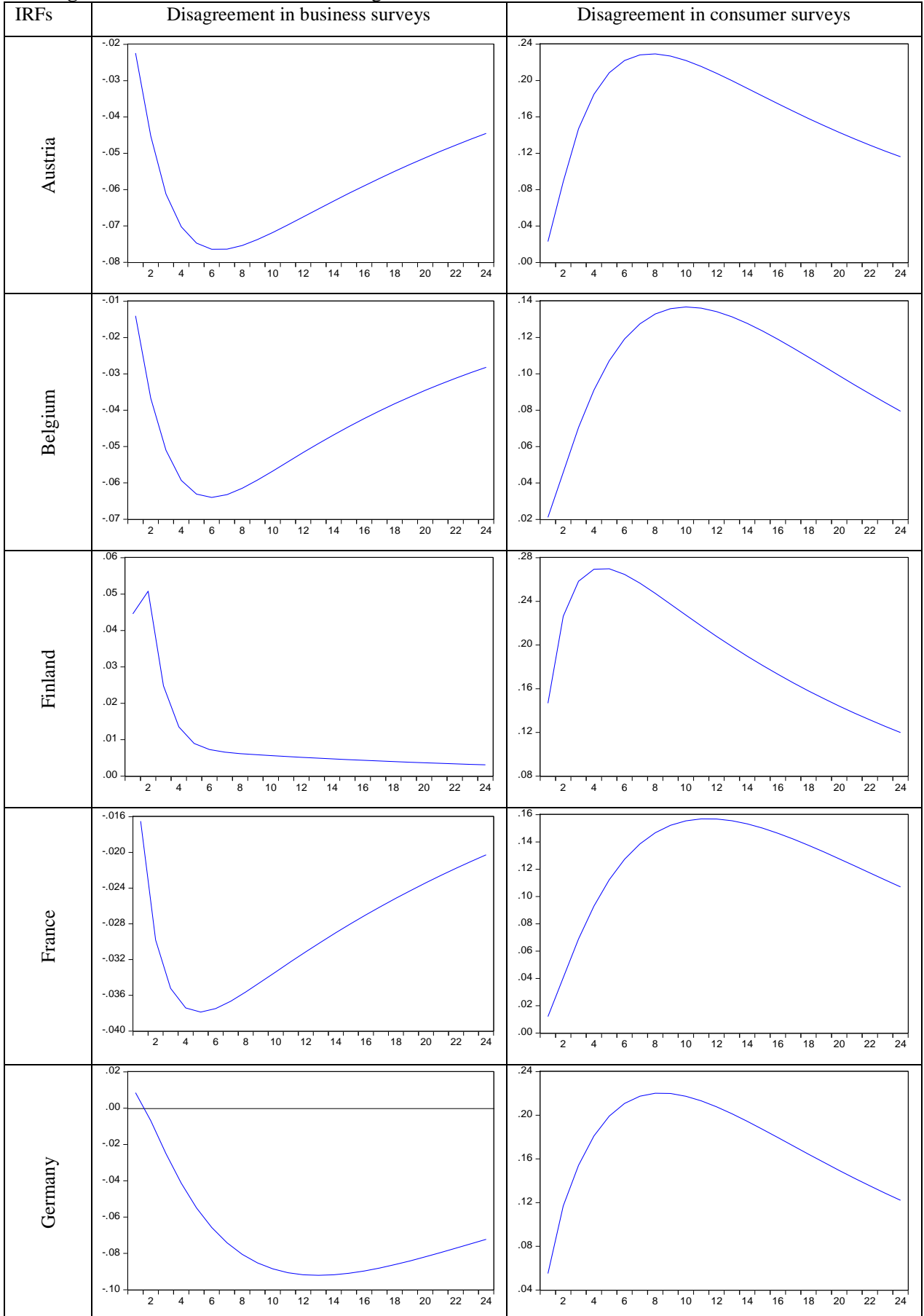

Notes: 24-month forecast horizon.

**Fig. 6** (cont.1)**.** IRFs of GDP to shocks in disagreement

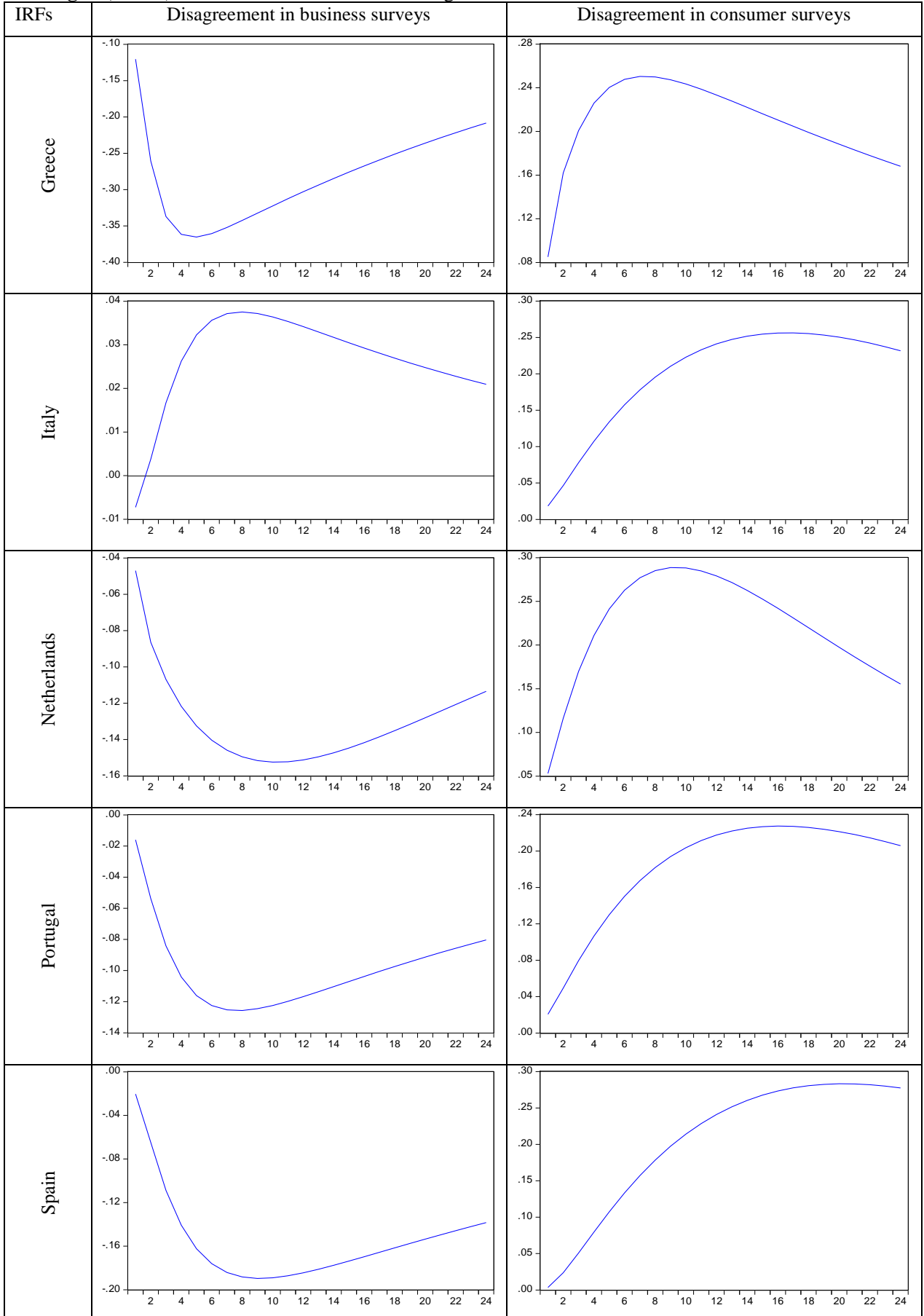

Notes: 24-month forecast horizon.

**Fig. 6** (cont.2)**.** IRFs of GDP to shocks in disagreement

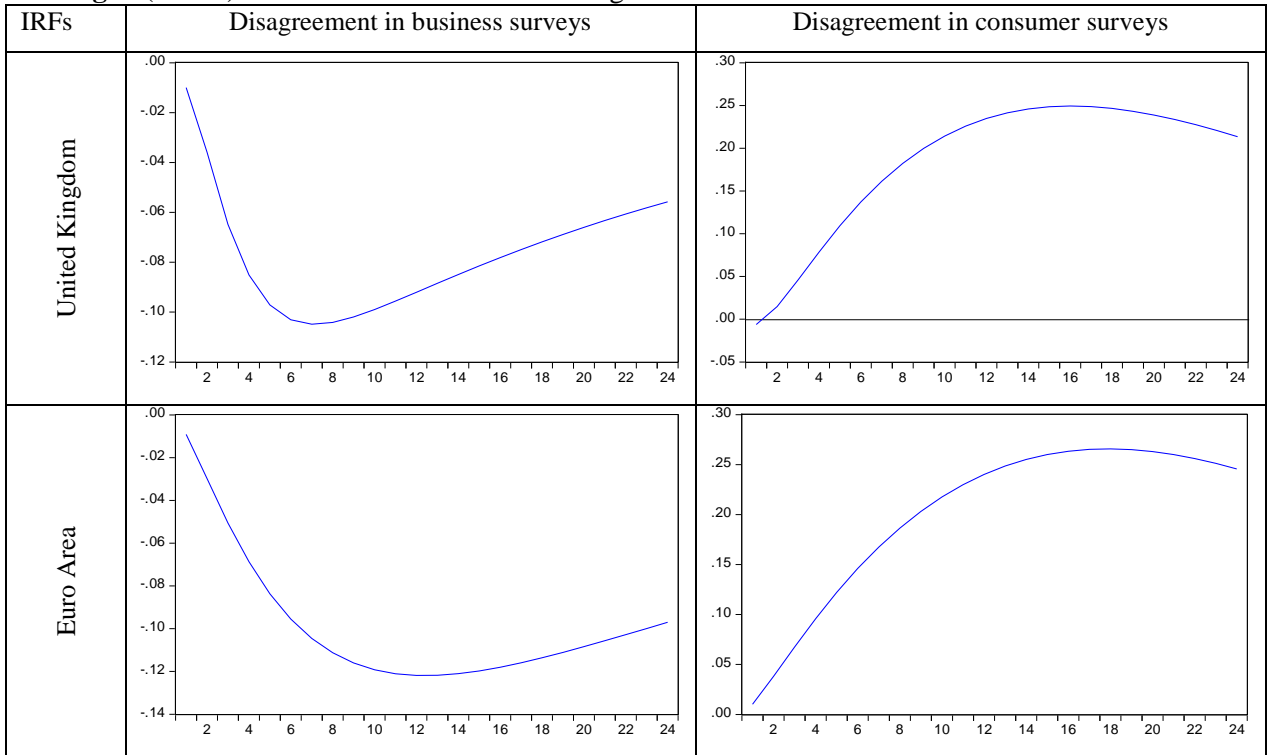

Notes: 24-month forecast horizon.

While Sahinoz and Cosar (2029) recently found that Turkish firms' and consumers' uncertainties coevolve, our analysis of the dynamic effects of shocks in the perception of economic uncertainty on GDP growth shows an asymmetric response between innovations in the disagreement in business surveys and in consumer surveys. A one standard deviation shock to DB leads to a fall in output growth. This result is in line with previous research (Alexopoulos and Cohen 2015; Cerda et al. 2018; Charles et al. 2018; Istiak and Serletis 2018; Meinen and Roehe 2017). On the contrary, a surprising result is that in most countries a one standard deviation shock to *DC* leads to an increase in output growth. This finding is partially in line with the results obtained by Morikawa (2019), who analysed the uncertainty of production forecasts and obtained heterogeneous forecast errors among individual manufactures and sectors. In this sense, Henzel and Rengel (2017) and Claveria (2020) found that different dimensions of uncertainty have diverse effects on aggregate fluctuations of the economy. Caggiano et al. (2017) and Netšunajev and Glass (2017) found evidence that unanticipated increases in uncertainty negatively affected the evolution of unemployment in the United States (US) and the EA.

We want to note that some of the obtained results may be conditioned by the setup of the analysis. As recently pointed out by Carriero et al. (2018), the fact that uncertainty measures are not fully embedded in the econometric models at the estimation stage might cause measurement errors in the regressors and lead to an endogeneity bias. Additional potential biases may also arise from the omission of variables due to restricted information sets in country-specific analysis. Some authors have circumvented this issue by assessing uncertainty shocks in a multi-economy context (Crespo et al. 2017; Mumtaz and Theodoridis 2017). Rossi and Sekhposyan (2015, 2017) introduced an index that allows to compute country-level contributions and helps to analyse the heterogeneity of uncertainty across countries. Other authors have included additional financial variables (Alessandri and Mumtaz 2019; Caldara et al. 2016; Gilchrist et al. 2014).

## 4. Conclusion

This study analyses the effect on economic growth of shocks in the perception of uncertainty of firms and consumers. We use qualitative data about the expected direction of change in economic activity, prices, and employment to proxy different dimensions of uncertainty which we then use to construct aggregate disagreement indicators for both businesses and consumers. Agents' perception of uncertainty is gauged by a geometric indicator of disagreement in survey expectations. First, we compare the level of disagreement between business and consumer surveys in eleven European countries and the Euro Area, and find that the average degree of consumer disagreement is greater than that of manufacturers.

Second, the dynamic relationship between innovations in perceived economic uncertainty and economic growth is assessed by estimating the impulse response functions using a Bayesian vector autoregressive framework. The obtained results differ markedly between disagreement in business and in consumer surveys. On the one hand, we find that shocks to consumer discrepancy tend to be of greater magnitude and duration than those to manufacturer discrepancy. On the other hand, we find that shocks to business discrepancy lead to a decrease in economic activity, as opposed to shocks to consumer economic discrepancy. This finding is of special relevance to researchers when using cross-sectional dispersion of survey-based expectations, since the effects of shocks to agents' perception of uncertainty on economic aggregates are shown to be dependent on the type of agent.

Finally, we want to note some of the limitations of the present study. On the one hand, it should be highlighted that the findings of this research may be conditioned by several biases derived from the exogenous measurement of uncertainty and the omission of variables. On the other hand, we want to point out the differences in the nature of the questions between business and consumer surveys, in the sense that the questions in the business survey always refer to specific factors of the company, while some of the questions from the consumer survey refer to the general development of economic activity. Regarding future lines of research, one aspect is the extension of the analysis to other variables included in the surveys, such as order-book levels, exports or savings, and to other surveys. Other aspects to explore are the extension of the methodological framework, using other measures of disagreement, applying new developments in VAR analysis.

**References**

Alessandri, P., and Mumtaz, H. (2019). Financial regimes and uncertainty shocks. *Journal of Monetary Economics*, 101, 31–46.
Alexopoulos, M., and Cohen, J. (2015). The power of print: Uncertainty shocks, markets, and the economy. *International Review of Economics and Finance*, 40, 8–28.
Atalla, T., Joutz, F., and Pierru, A. (2016). Does disagreement among oil price forecasters reflect volatility? Evidence form the ECB surveys. *International Journal of Forecasting*, 32(4), 1178–1192.
Bachmann, R., and Bayer, C. (2013). 'Wait-and-See' business cycles. *Journal of Monetary Economics*, 60(6), 704–719.
Bachmann, R., Elstner, S., and Sims, E. R. (2013). Uncertainty and economic activity: Evidence from business survey data. *American Economic Journal: Macroeconomics*, 5(2), 217–249.
Baker, S. R., Bloom, N., and Davis, S. J. (2016). Measuring economic policy uncertainty. *Quarterly Journal of Economics*, 131(4), 1593–1636.
Balcilar, M., Bekiros, S., and Gupta, R. (2017). The role of news-based uncertainty indices in predicting oil markets: a hybrid nonparametric quantile causality method. *Empirical Economics*, 53(3), 879–889.
Basu, S, and Bundick, B. (2017). Uncertainty shocks in a model of effective demand. *Econometrica*, 85(3), 937–958.
Bekaert, G., Hoerova, M., and Lo Duca, M. (2013). Risk, uncertainty and monetary policy. *Journal of Monetary Economics*, 60(7), 771–788.
Binder, C. (2017). Measuring uncertainty based on rounding: New method and application to inflation expectations. *Journal of Monetary Economics*, 90, 1–12.
Binding, G., and Dibiasi, A. (2017). Exchange rate uncertainty and firm investment plans evidence from Swiss survey data. *Journal of Macroeconomics*, 51, 1–27.
Bloom, N. (2009). The impact of uncertainty shocks. *Econometrica*, 77(3), 623–685.
Caggiano, G., Castelnuovo, E., and Figueres, J. M. (2017). Economic policy uncertainty and unemployment in the United States: A nonlinear approach. *Economics Letters*, 151, 31–34.
Caldara, D., Fuentes-Albero, C., Gilchrist, S., and Zakrajšek, E. (2016). The macroeconomic impact of financial and uncertainty shocks. *European Economic Review*, 88, 185–207.

Carriero, A., Clark, T. E., and Marcellino, M. (2018). Measuring uncertainty and its impact on the economy. *Review of Economics and Statistics*, 100(5), 799–815.
Cerda, R., Silva, A., and Valente, J. T. (2018). Impact of economic uncertainty in a small open economy: The case of Chile. *Applied Economics*, 50(26), 2894–2908.
Charles, A., Darné, O., and Tripier, F. (2018). Uncertainty and the macroeconomy: Evidence from an uncertainty composite indicator. *Applied Economics*, 50(10), 1093–1107.
Chuliá, H., Guillén, M., and Uribe, J. M. (2017). Measuring uncertainty in the stock market. *International Review of Economics and Finance*, 48, 18–33.
Claveria, O. (2018). A new metric of consensus for Likert scales. *AQR Working Papers* 2018/10. University of Barcelona, Regional Quantitative Analysis Group.
Claveria, O. (2019). A new consensus-based unemployment indicator. *Applied Economics Letters*, 26 (10), 812–817.
Claveria, O. (2019). Forecasting the unemployment rate using the degree of agreement in consumer unemployment expectations. *Journal for Labour Market Research*, 53(3), 1–10.
Claveria, O. (2020). Uncertainty indicators based on expectations of business and consumer surveys. *Empirica*. Forthcoming.
Claveria, O., Monte, E., Torra, S. (2019). Economic uncertainty: A geometric indicator of discrepancy among experts' expectations. *Social Indicators Research*, 143(1), 95–114.
Clements, M., and Galvão, A. B. (2017). Model and survey estimates of the term structure of US macroeconomic uncertainty. *International Journal of Forecasting*, 33(3), 591–604.
Crespo, J., Huber, F., and Onorante, L. (2017). The macroeconomic effects of international uncertainty shocks. WU Working Paper Series 245.
Dovern, J. (2015). A multivariate analysis of forecast disagreement: Confronting models of disagreement with survey data. *European Economic Review*, 80, 1–12.
Gilchrist, S., Sim, J. W., and Zakrajšek, E. (2014). Uncertainty, financial frictions, and investment dynamics. NBER Working Paper 20038.
Girardi, A., and Reuter, A. (2017). New uncertainty measures for the euro area using survey data. *Oxford Economic Papers*, 69(1), 278–300.
Glass, K., and Fritsche, U. (2014). Real-time information content of macroeconomic data and uncertainty: An application to the Euro area. *DEP (Socioeconomics) Discussion Papers*, Macroeconomics and Finance Series 6/2014, University of Hamburg.
Glocker, C., and Hölzl, W. (2019). Assessing the economic content of direct and indirect business uncertainty measures. WIFO Working Paper 576/2019.
Gupta, R., Lau, C. K. M., and Wohar, M. E. (2019). The impact of US uncertainty on the Euro area in good and bad times: Evidence from a quantile structural vector autoregressive model. *Empirica*, 46(2), 353–368.
Hailemariam, A., and Smyth, R. (2019). What drives volatility in natural gas prices?. *Energy Economics*. 80, 731–742.
Henzel, S., and Rengel, M. (2017). Dimensions of macroeconomic uncertainty: A common factor analysis. *Economic Inquiry*, 55(2), 843–877.
Istiak, K., and Serletis, A. (2018). Economic policy uncertainty and real output: Evidence from the G7 countries. *Applied Economics*, 50(39), 4222–4233.
Jurado, K., Ludvigson, S., and Ng, S. (2015). Measuring uncertainty. *American Economic Review*, 105(3), 1177–216.
Krüger, F., and Nolte, Ingmar. (2016). Disagreement versus uncertainty: Evidence from distribution forecasts. *Journal of Banking & Finance*, Supplement, 72, S172–S186.
Lahiri, K., and Sheng, X. (2010). Measuring forecast uncertainty by disagreement: The missing link. *Journal of Applied Econometrics*, 25(4), 514–38.
Mankiw, N. G., Reis, R., and Wolfers, J. (2004). Disagreement about inflation expectations. In M. Gertler and K. Rogoff (Eds.), *NBER Macroeconomics Annual 2003* (pp. 209–248). Cambridge, MA: MIT Press.
Meinen, P., and Roehe, O. (2017). On measuring uncertainty and its impact on investment: Cross-country evidence from the euro area. *European Economic Review*, 92, 161–179.
Mitchell, J., Mouratidis, K., and Weale, M. (2007). Uncertainty in UK manufacturing: Evidence from qualitative survey data. *Economics Letters*, 94(2), 245–135.

Mokinski, F., Sheng, X., and Yang, J. (2015). Measuring disagreement in qualitative expectations. *Journal of Forecasting*, 34(5), 405–426.

Morikawa, M. (2019). Uncertainty over production forecasts: An empirical analysis using monthly quantitative survey data. *Journal of Macroeconomics*, 60, 163–179.

Mumtaz, H., and Theodoridis, K. (2017). Common and country specific economic uncertainty. *Journal of International Economics*, 105, 205 – 216.

Mumtaz, H., and Musso, A. (2019). The evolving impact of global, region-specific and country-specific uncertainty. *Journal of Business & Economic Statistics*. Forthcoming.

Netšunajev, A., and Glass, K. (2017). Uncertainty and employment dynamics in the euro area and the US. *Journal of Macroeconomics*, 51, 48–62.

Oinonen, S., and Paloviita, M. (2017). How informative are aggregated inflation expectations? Evidence from the ECB Survey of Professional Forecasters. *Journal of Business Cycle Research*, 13(2), 139–163.

Paloviita, M., and Viren, M. (2014). Inflation and output growth uncertainty in individual survey expectations. *Empirica*, 41(1), 69–81.

Rossi, B., and Sekhposyan, T. (2015). Macroeconomic uncertainty indices based on nowcast and forecast error distributions. *American Economic Review*, 105(5), 650–655.

Rossi, B., and Sekhposyan, T. (2017). Macroeconomic uncertainty indices for the euro area and its individual member countries. *Empirical Economics*, 53(1), 41–62.

Sahinoz, S., and Cosar, E. E. (2020). Quantifying uncertainty and identifying its impacts on the Turkish economy. *Empirica*, 47(2), 365–387.

Yıldırım-Karaman, S. (2017). Uncertainty shocks, central bank, characteristics and business cycles. *Economic Systems*, 41(3), 379–388.

Zarnowitz, V., and Lambros, L. A. (1987). Consensus and uncertainty in economic prediction. *Journal of Political Economy*, 95(3), 591–621.

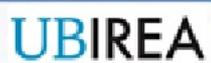

Institut de Recerca en Economia Aplicada Regional i Públic
*Research Institute of Applied Economics*

**WEBSITE:** www.ub-irea.com • **CONTACT:** irea@ub.edu

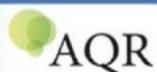

Grup de Recerca Anàlisi Quantitativa Regional
*Regional Quantitative Analysis Research Group*

**WEBSITE:** www.ub.edu/aqr/ • **CONTACT:** aqr@ub.edu

**Universitat de Barcelona**
Av. Diagonal, 690 • 08034 Barcelona